\documentstyle{elsart}
\begin{document}

\hfill LMU-TPW-97-9

\begin{frontmatter}
\title{A ``Gaussian'' Approach to Computing Supersymmetric Effective Actions}

\author{I.N. McArthur\thanksref{UWA} \thanksref{AVH}} and
\author{T.D. Gargett\thanksref{UWA}}
\address{Sektion Physik, LS Wess, Universit\"at M\"unchen, Theresienstrasse
37,
D-80333 M\"unchen, Germany.}

\thanks[UWA]{Permanent address: Physics Department, The University of Western
Australia, Nedlands, W.A. 6907, Australia.}
\thanks[AVH]{ Alexander von Humboldt  Research Fellow}
\begin{abstract}
For nonsupersymmetric theories, the one-loop effective action can be computed
via  zeta function regularization in terms of the functional trace of the heat
kernel associated with the operator which appears in the quadratic part of the
action. A method is developed for computing this  functional trace by
exploiting its similarity to  a Gaussian integral. The procedure is
extended to
superspace, where it is used to compute  the low
energy effective action obtained by integrating out massive scalar
supermultiplets in the presence of a supersymmetric Yang-Mills background.
\end{abstract}
\end{frontmatter}

\section{Introduction}
Low energy effective actions are important in physical situations in
which one is interested in phenomena at an energy scale which is
small compared to the masses of some of the fundamental degrees of
freedom in the theory. Although these heavy degrees of freedom can only
occur as virtual states at the energy scale of interest, they still
produce observable effects, which are summarized in a low energy
effective action for the ``light" degrees of freedom. The low energy
effective action is determined by ``integrating out" the heavy degrees
of freedom.

 Recently there has been  renewed interest in perturbative computations of
 effective actions in supersymmetric Yang-Mills theories
\cite{henningson,dewit,pickering,grisaru}. This interest was
stimulated by the work of Seiberg and Witten \cite{SeibergWitten}, who
deduced the full
nonperturbative low energy effective action for an N=2 supersymmetric
Yang-Mills theory. The perturbative computations have focused on the
corrections to the low energy  K\"ahler potential for the  scalar
supermultiplets in the N=1 superfield formulation of N=2
supersymmetric Yang-Mills theory. The effective K\"ahler potential
was first  considered in superfield form in the work of
Buchbinder et al \cite{buchbinder1,buchbinder2}.

In this paper, we introduce a new technique for the perturbative
calculation of
low energy effective actions, and then illustrate  it in superspace by
determining the low energy effective action  obtained by
integrating out massive N=1  scalar multiplets in the presence of an N=1
supersymmetric
Yang-Mills background. This is a case not treated fully in the earlier
calculations quoted above. The approach is from the point of view of
heat kernels and zeta functional regularization, as opposed to the
supergraph calculations in \cite{dewit,pickering,grisaru}.
Buchbinder et al \cite{buchbinder1,buchbinder2} have also developed functional
techniques for computing low energy effective actions in superspace,
but
our approach differs significantly from theirs. Also, although
zeta function
regularization is only really useful in the computation of one-loop effective
actions, there are
nonrenormalization theorems in N=2 supersymmetric Yang-Mills theories
 which ensure the absence of higher loop corrections
to the effective action, so this is not a disadvantage if one is
ultimately interested in these theories.

The plan of the paper is as follows.  We begin with  a brief summary of an
approach to the computation of nonsupersymmetric low-energy effective actions
which relies on the similarity  of the functional trace of the heat kernel
to a
Gaussian integral.  The technique  is then applied in superspace to determine
the low energy effective action  obtained by
integrating out massive N=1  scalar multiplets coupled to an N=1
supersymmetric
Yang-Mills background. The paper ends with a short discussion.\\

\section{ The Nonsupersymmetric Case}
An efficient way to generate the one-loop effective action  is via zeta
function
regularization. We consider for definiteness a massive scalar field in the
presence of a Yang-Mills background with  Euclidean action $$S[A, \phi] =
\frac12\int d^4x \, \phi^{\dagger}  (-D_aD_a + m^2) \phi,$$ where $D_a$ is the
covariant derivative in the   representation $R$ of the gauge group to
which the
scalar fields belong, with $[D_a,D_b] = F_{ab}.$  The one loop effective
action
for the Yang-Mills fields  obtained by ``integrating out'' the scalar
fields is
$$\Gamma[A] =- \ln  \int [d\phi] e^{-S[A,\phi]} =  \ln \det (-D_aD_a + m^2). $$
 Using zeta function regularization, this  is just $ -  \zeta'(0), $ where the
zeta function is defined by
$$ \zeta(s) = \frac{1}{\Gamma(s)} \int_0^{\infty} ds \, t^{s-1} e^{-m^2
t/\mu^2}\,  Tr {\tilde K}(t/\mu^2). $$
In this expression, $Tr$ is a trace over gauge indices and ${\tilde K}(t)$ is
the functional trace of the heat kernel for the operator $D_aD_a,$
$$ {\tilde K}(t) = \int d^4x \, \lim_{x \rightarrow x'} e^{tD_aD_a}\,
\delta^{(4)}(x,x') \equiv  \int d^4x K(t). $$
Since $t$ in $K(t)$ has dimensions of inverse mass squared, the usual
parameter
$\mu$ with dimensions of mass (representing the renormalization point)  has
been
introduced into the zeta function to make $t$ dimesionless.

If we introduce a plane wave basis for the delta function, $K(t)$ takes the
form
 \begin{eqnarray*}
K(t) & =&   \lim_{x \rightarrow x'} \int \frac{d^4k}{(2\pi)^4} \, e^{i
k.(x-x')}
\, \left(  e^{-ik.(x-x')} \, e^{t D_a D_a} \, e^{ik.(x-x')} \right) \\
&=& \int  \frac{d^4k}{(2\pi)^4} \, e^{t X_a X_a }
\end{eqnarray*}
where $X_a = D_a + i k_a.$ It then follows that $K(t)$ satisfies the
differential equation
\begin{equation}
\frac{dK(t)}{dt} = K_{aa}(t),
\label{deq}
\end{equation}
 where the tensors $K_{a_1 \, \cdots \, a_n}(t)$ are defined by
$$K_{a_1 \, \cdots \, a_n}(t) = \int \frac{d^4k}{(2\pi)^4} \, X_{a_1} \,
\cdots
\, X_{a_n} \, e^{t X.X}. $$
To solve the differential equation, it is necessary to obtain an expression
for
$K_{ab}(t)$ in terms of $K(t).$ The approach we take is to  use the identity
$$ 0 = \int \frac{d^4k}{(2\pi)^4} \, \frac{\partial}{\partial k^{a_m}} \,
\left(
X_{a_1} \, \cdots \, X_{a_{m-1}} \, e^{tX.X} \right)  $$
to study the properties of these tensors (the boundary term in the integral
vanishes because of the $e^{-k^2}$ factor  in the integrand). This is the same
method that can be used to determine the moments $\int
\frac{d^4k}{(2\pi)^4} \,
k_{a_1} \cdots k_{a_n} e^{-k^2}$ of an ordinary Gaussian in terms of the
Gaussian itself.

In particular, applying it in the case $m=2,$
\begin{eqnarray}
 0 &= & i \delta_{ab } K(t) +  \int \frac{d^4k}{(2\pi)^4} \, X_{a} \,
\frac{\partial}{\partial k^{b}} \,  e^{tX.X} \nonumber \\
 &= & i \delta_{ab} K(t) + 2 i t \int \frac{d^4k}{(2\pi)^4} \, X_{a}  \left(
\int_0^1 ds \, e^{-st X.X} X_{b} \, e^{stX.X}\right) e^{tX.X}.
\label{Kuv}
\end{eqnarray}
Because  $\int_0^1 ds \, e^{-st X.X} X_{b}\,  e^{stX.X} =   X_b + \,
\cdots, $
this identity yields an expression for $K_{ab}(t) $ in terms of $K(t)$  which
can then be used to solve the linear  differential equation (\ref{deq}).

The quantity
\begin{equation}
\int_0^1 ds \, e^{-st X.X} \, X_{b} \, e^{stX.X} = \sum_{n=0}^{\infty}
\frac{t^n}{(n+1)!} \, ad^{(n)}(X.X)(X_b)
\label{ints}
\end{equation}
cannot be evaluated exactly and so must be approximated in some way. In the
case
of a computation of the low energy effective action, one is interested in the
piece of the effective action which contains no covariant derivatives of the
field strength of the Yang-Mills background.  This is the covariant
generalization of the low energy effective action for the $U(1)$ case, where a
constant field strength  is equivalent to the long wavelength limit for the
background electromagnetic field.       A covariantly constant background
corresponds to setting  $(D_{a} F_{bc})  = [X_{a}, [X_{b}, X_{c}]] $ to
zero, so
that only the commutators $[X_a,X_b ] = F_{ab}$ need be retained in
evaluating
(\ref{ints}).  In this case, it is easy to show that $  ad^{(n)}(X.X)(X_{b}) =
(-2)^n (F^n)_{b c} X_{c}, $ so that
$$  \int_0^1 ds \, e^{-st X.X} X_{b} \, e^{stX.X} = B_{bc}(t) \,  X_{c} $$
where
$$ B_{bc}(t) =  \left[ \frac{e^{-2tF}-{\bf 1}}{-2tF} \right]_{bc}.    $$
Note that the power series expansion for the matrix $B_{bc}(t) $ begins with
$\delta_{bc}$  and only involves positive powers of $F_{ab}.$ As a result, the
inverse matrix $B^{-1}_{bc}(t) =   \left[ \frac{-2tF}{e^{-2tF}-{\bf
1}}\right]_{bc}$ exists and does not require the inverse of $F_{ab}$ to be
determined.
Inserting this result into (\ref{Kuv}), and using the fact that $B_{bc}(t)$
commutes with $X_a$ to the order that we are working,
one  finds
$$ 0 = \delta_{ab} \, K(t) + 2 t B_{bc}(t) K_{ac}(t).$$
It follows that
$$ K_{ab} =  \left[ \frac{F}{e^{-2tF}-{\bf 1}} \right]_{ba}. $$
Thus (\ref{deq}) becomes
$$
\frac{d K(t)}{dt} =  tr  \left[ \frac{F}{e^{-2 t F}-{\bf 1}}\right] K(t),
$$
where the trace $tr$ is over spacetime indices and {\em not} gauge indices;
the
kernel is still a matrix with respect to its gauge indices. Noting that $ tr
\left[ \frac{F}{e^{-2 t F}-{\bf 1}}\right]$  $ = tr  \left[
\frac{Fe^{2tF}}{{\bf
1} -e^{2tF}}\right] = -\frac12  tr \left[C^{-1} ({\bf 1}-e^{2tF})^{-1}
\frac{d}{dt} ({\bf 1}-e^{2tF}) C\right],$ with $C$ a matrix independent of
$t$,
and using the boundary condition that $K(t)$ reduces to the ordinary Gaussian
$\int \frac{d^4k}{(2\pi)^4} e^{-k^2}$ in the limit $F_{ab} \rightarrow 0,$
one finds  the standard result \cite{schwinger,brown,reuter,schubert} for the
functional trace of the heat kernel:
\begin{equation}
K(t) = \frac{1}{4 \pi^2}  \det \left[\frac{ {\bf 1}-e^{2tF}
}{F}\right]^{-\frac12} = \frac{1}{16 \pi^2 t^2} \, \det \left[ \frac{t
F}{\sinh
tF} \right]^{\frac12}.
\label{le}
\end{equation}

This technique is readily generalises to quantum fields of different spin,
or to
the inclusion of a potential for the scalar fields. For later use, we note
that
for a Dirac spinor in a Yang-Mills background, the kernel for the Laplace-type
operator given by the square of the Dirac operator is  \cite{schwinger,brown}
\begin{equation}
K_{\frac12}(t) =  tr \left(e^{- t \Sigma_{ab}F_{ab}} \right)  \, K(t),
\label{spin12}
\end{equation}
where  $\Sigma_{ab} = \frac{1}{4}[\gamma_a,\gamma_b],$ the trace $tr$ is over
the spinor indices on the gamma matrices and $K(t)$ is the spin zero kernel
(\ref{le}).

The above considerations were motivated to a certain extent by the work of
Avramidi \cite{avramidi}, who computed
$$\int \frac{d^4k}{(2\pi)^4} \, \sqrt{\gamma} \, e^{-t( \gamma_{ab}k_ak_b -
2
k_a D_a)},$$
where $\gamma_{ab}$ is a metric in $k$ space, and $\gamma = \det \gamma_{ab}.$
The result is of the form $f(t,\gamma,F) \, e^{t g_{ab}(t)D_aD_b},$ where
$g_{ab}$ is a functional of $\gamma_{ab}$ and $F_{ab}.$ By choosing
$\gamma_{ab}$ so that $g_{ab} = \delta_{ab},$ it is possible to obtain an
expression for $e^{tD_aD_a}$ which can be used to compute the heat kernel and
hence its functional trace. It is not immediately clear how this could be
applied to superspace calculations.

Another  common approach to computing effective actions is to determine the
Green's function  for the quantum  fields  in the presence of the  background;
this is then  used to compute the functional trace of the heat kernel. In the
above approach,  we compute the functional trace of the heat kernel directly
without the need for this intermediate step. This is an advantage when it
comes
to computing effective actions in supersymmetric theories, as there are many
different Green's functions. In \cite{buchbinder1}, where the effective
K\"ahler
potential for the Wess-Zumino model is computed,  Buchbinder et al
expressed all
the Green's functions in terms of a single one, thereby providing some
simplification. In the next section, we show that it is possible to extend the
techniques developed above to calculations of effective actions for
supersymmetric theories in superspace.  To illustrate this, we  compute the
supersymmetric analogue of the results  (\ref{le}) and (\ref{spin12}).

\section{  Massive Scalar Multiplet  in Gauge Superfield Background}
Here, we will be  concerned with the computation of the one-loop low energy
effective action  for a gauge supermultiplet which results from integrating
out
massive scalar multiplets coupled to the gauge background.
  This is the superspace analogue of the nonsupersymmetric theory treated in
\S2.  The superspace action for the scalar supermultiplet $\Phi(x,\theta)$
transforming in some (real) representation $R$ of the gauge group $G$ is
$$S = \int d^4x \, d^2\theta \, d^2\bar{\theta} \, \, \bar{\Phi} \, e^{-V}
\Phi
+ \int d^4x \,d^2\theta \, \frac{m}{2} \Phi^2 + \int d^4x \,d^2\bar{\theta} \,
\frac{m}{2} \bar{ \Phi}^2. $$
Gauge indices are suppressed, and the superspace conventions of  of Wess and
Bagger \cite{Wess} have been adopted, except that the metric has been Wick
rotated to be Euclidean. Because the quantum superfields $\Phi$ are subject to
the chirality constraint $\bar{D}_{\dot{\alpha}} \Phi = 0,$ the effective
action
cannot simply be formed as the logarithm of the superdeterminant of the
operator
appearing in the quadratic part of the action. Rather, it is first
necessary to
write the action in terms of unconstrained superfields. To this purpose we
introduce complex scalar superfields $\Psi$ and solve the constraint by
expressing $\Phi = \bar{D}^2  \Psi;$ the superfields transform under the
action
of the gauge group in the same way as  $\Phi$. In terms of the unconstrained
fields, the action  can be expressed in the full superspace as
$$ S = 8  \int d^4x \, d^2\theta \, d^2\bar{\theta} \left( \begin{array}{ll}
\Psi, & \Psi^{\dagger}  \end{array} \right)   \left(
\begin{array}{ll}  \frac{1}{16} \bar{D}^2e^V D^2 e^{-V} & -\frac{m}{4}
\bar{D}^2e^V \\ -\frac{m}{4} D^2e^{-V} & \frac{1}{16} D^2e^{-V} \bar{D}^2 e^V
\end{array} \right)  \left( \begin{array}{c} e^{V}  \Psi^{\dagger}  \\
e^{-V}
\Psi   \end{array} \right) .$$
So the effective action is $\Gamma_{eff}[V] = - \frac12 \ln {\rm sdet}
\Delta, $
where $\Delta $ is the superspace operator in the unconstrained action above
\cite{buchbinder1}. The appropriate zeta function is thus
$ \zeta(s) = \frac{1}{\Gamma(s)} \int_0^{\infty} dt \, t^{s-1}  \int d^8Z
\, Tr
 K(t/\mu^2), $
where $Tr$ denotes the trace over gauge indices, $\mu$ is the renormalization
point  and
$$
K(t) = tr  \lim_{Z \rightarrow Z'} \exp t  \left(  \begin{array}{ll}
\frac{1}{16} \bar{D}^2e^V D^2e^{-V} & -\frac{m}{4} \bar{D}^2e^V \\
-\frac{m}{4}
D^2e^{-V} & \frac{1}{16} D^2e^{-V} \bar{D}^2 e^V \end{array} \right) \,
\delta^{(8)}(Z,Z'). $$
Here, the trace $tr$  is over  the $ 2\times 2 $ matrices, and $\int d^8Z$
and
$ \delta^{(8)}(Z,Z') $ denote the integration measure and  the delta
function on
the full superspace.
Applying the Baker-Campbell-Hausdorff formula, the  terms involving the
mass $m$
 can be placed in a separate exponential; performing the two dimensional trace
then projects out even powers of $m$, giving
\begin{eqnarray*}
 K(t) & =  &  \sum_{n=0}^{\infty}  \frac{(mt)^{2n}}{(2n)!}\, \lim_{Z
\rightarrow
Z'}\,
\biggl(      \left( \frac{1}{16} \bar{D}^2e^VD^2e^{-V} \right)^n \,
e^{\frac{t}{16} \bar{D}^2e^VD^2e^{-V}}\delta^{(8)}(Z,Z') \\
& + &
\left(\frac{1}{16} D^2e^{-V}\bar{D}^2e^{V} \right)^n \,   e^{\frac{t}{16}
D^2e^{-V}\bar{D}^2e^{V}} \delta^{(8)}(Z,Z')  \biggr) \\
& = &  \sum_{n=0}^{\infty}  \frac{(mt)^{2n}}{(2n)!}\,  \frac{d^n}{dt^n} \,
\lim_{Z \rightarrow Z'} \biggl( e^{\frac{t}{16}
\bar{D}^2e^VD^2e^{-V}}\delta^{(8)}(Z,Z') \\
& +  &   e^{\frac{t}{16} D^2e^{-V}\bar{D}^2e^{V}} \delta^{(8)}(Z,Z')  \biggr).
\end{eqnarray*}
 Using $\int d^8Z  = \int d^4x \int d^2 \theta  ( - \frac14 ) \bar{D}^2$ and
$\int d^8Z  = \int d^4x \int d^2 \bar{ \theta}  ( - \frac14 ) D^2,$
factors of
$( - \frac14 ) \bar{D}^2$ and $ (- \frac14 ) D^2$ can be extracted from the
exponentials to act  on the full superspace delta function to convert the zeta
function to the ``chiral'' form
\begin{eqnarray}
\zeta(s) &=&  \frac{1}{\Gamma(s)} \int_0^{\infty} dt \, t^{s-1} \,
\sum_{n=0}^{\infty} \frac{1}{(2n)!} \left( \frac{mt}{\mu^2}\right)^{2n}\,
\mu^{2n}  \frac{d^n}{dt^n}\, Tr \,\biggl( \int d^4x \int d^2 \theta \,
K_L(t/\mu^2) \nonumber  \\
&  + &  \int d^4x \int d^2 \bar{\theta}  \, K_R(t/\mu^2) \biggr) ,
\label{zetasum}
\end{eqnarray} where the chiral kernels are
\begin{eqnarray}
K_L(t) &=& \lim_{Z \rightarrow Z'} e^{\frac{t}{16}  \bar{D}^2e^VD^2e^{-V}}
\delta^{(4)}(x,x') \delta^{(2)}(\theta, \theta'), \nonumber \\
K_R(t) &=& \lim_{Z \rightarrow Z'} e^{\frac{t}{16}  D^2e^{-V}\bar{D}^2e^{V}}
\delta^{(4)}(x,x') \delta^{(2)}(\bar{\theta}, \bar{\theta}').
\label{chiralkernels}
\end{eqnarray}

The mass dependence in (\ref{zetasum}) involves derivatives of the chiral
kernels. It is convenient to remove these derivatives by repeated
integration by
parts. The boundary terms at $t= \infty $ vanish as the kernels vanish in this
limit (this can be checked explicitly using the result  (18) for the kernel);
the boundary term at $t=0$  involves a factor $t^{n+s+1}$ and also vanishes as
we are only interested in $\zeta(s)$ for $s$ in a small neighbourhood of
$s=0.$
The result can be expressed in the form
\begin{eqnarray*}
\zeta(s) &=&  \frac{1}{\Gamma(s)} \int_0^{\infty} dt \, t^{s-1} \, \left[
\sum_{n=0}^{\infty}  \frac{1}{(2n)!} \left(-\frac{m^2t}{\mu^2}\right)^n\,
\frac{\Gamma (2n+s)}{\Gamma (n+s) }\right ]\\ & &  Tr \,\biggl( \int d^4x \int
d^2 \theta \,  K_L(t/\mu^2)
 +    \int d^4x \int d^2 \bar{\theta}  \, K_R(t/\mu^2) \biggr) .
\end{eqnarray*}
The sum in square brackets is the generalized hypergeometric function
$$ {}_2F_2\left[ \frac{s}{2} + \frac12,\frac{s}{2};
\frac12,s;-\frac{m^2t}{\mu^2} \right]. $$
 It will be seen later that the chiral kernels $ \int d^4x \int d^2 \theta \,
K_L(t)   +   \int d^4x \int d^2 \bar{\theta}  \, K_R(t)$ have  a power series
expansion in $t$ of the form $\sum_{n=0}^{\infty} V_n t^n,$ where $V_n$
represents an effective vertex for the interaction of $n+2$ particles. So,
with
a rescaling of $t$,
$$ \zeta(s) = \sum_{n=0}^{\infty} \frac{V_n}{ m^2}
\left(\frac{m^2}{\mu^2}\right)^{-s}\, \frac{1}{\Gamma(s)} \int_{0}^{\infty} dt
\, t^{n+s-1}  {}_2F_2\left[ \frac{s}{2}+\frac12 ,\frac{s}{2}; \frac12,s ;-t
\right].  $$
Although we have not been able to do the integral (which is the Mellin
transform
of a generalized hypergeometric function)  explicitly, it is possible to
deduce
the form of the zeta function for $s $ near zero using the fact that $
{}_2F_2\left[ \frac{s}{2}+\frac12 ,\frac{s}{2}; \frac12,s;-t \right]
\rightarrow
\frac12 e^{-t}$ as $s \rightarrow 0.$ Thus for
$ n\neq 0, $ the integral is regular in the limit $s \rightarrow 0,$  giving
$\frac12 \Gamma(n). $ This means that in calculating $\zeta'(0), $ the
derivative must
act on $\frac{1}{\Gamma (s)} = s + O(s^2)$ to eliminate the zero at $s=0.$ For
$n=0,$ the integral behaves like $\frac12 \Gamma(s) $ for small $s,$ and so
the
derivative acts on $ (m^2/\mu^2)^{-s}. $ The result is
\begin{equation}\zeta'(0) = \sum_{n=1}^{\infty}\frac12\, \Gamma(n)
\frac{V_n}{m^{2n}}  -
\frac12 \,V_0 \ln \frac{ m^2}{\mu^2}.   \label{seriesexp}
\end{equation}
Alternatively, if Schwinger proper time regularization is used, it is only
necessary to know the hypergeometric at $s=0;$ in this case there is a
divergence in the $n=0$ contribution to the effective action which must be
removed by hand, but the $n\geq 1$ contributions are as above.

It therefore remains to evaluate the chiral kernels; we do this for
$K_L(t),$ as
the calculation for $K_R(t)$ is identical except that  left chiral quantities
are  replaced by right chiral quantities. For the rest of the paper, $K_L(t)$
will simply be denoted $K(t).$ Acting on left chiral superfields, the
operator $
\frac{1}{16}  \bar{D}^2e^VD^2e^{-V}$ is equivalent to the Laplace-type
operator
${\cal D}_a {\cal D}_a + W^{\alpha} {\cal D}_{\alpha} + \frac12 ({\cal
D}^{\alpha}W_{\alpha}),$ where the ${\cal D}$ denote gauge covariant
superspace
derivatives in the left chiral basis:  ${\cal D}_{\alpha} = e^V D_{\alpha}
e^{-V},$ $\bar{{\cal D}}_{\dot{\alpha}} = \bar{ D}_{\dot{\alpha}},$ $ \{ {\cal
D}_{\alpha}, \bar{{\cal D}}_{\dot{\alpha}} \} =  -2 i (\sigma_a)_{\alpha
\dot{\alpha} } \,{\cal D}_a, $ and $W_{\alpha} = -\frac14 [\bar{{\cal
D}}_{\dot{\alpha}}, \{ \bar{{\cal D}}^{\dot{\alpha}}, {\cal D}_{\alpha} \} ].$
Thus the left chiral kernel contains a Laplace-type operator, as required
for a
well-defined heat kernel:
$$ K(t) = \lim_{Z \rightarrow Z'} e^{t({\cal D}_a {\cal D}_a + W^{\alpha}
{\cal
D}_{\alpha} + \frac12 ({\cal D}^{\alpha}W_{\alpha}))} \, \delta^{(4)}(x,x')
\delta^{(2)}(\theta, \theta'). $$
The  left chiral delta function has the representation
$$\frac14 \delta^{(4)}(x,x')  \delta^{(2)}(\theta, \theta') = \int
\frac{d^4k}{(2\pi)^4}\, e^{ik_a (x_a - x_a' - i \theta \sigma_a
\bar{\theta}' +
i \theta' \sigma_a \bar{\theta})}  \int d^2 \epsilon\,  e^{i
\epsilon^{\alpha}(\theta - \theta')_{\alpha}}, $$
where $\epsilon_{\alpha}$ is a Grassmann parameter which is the supersymmetric
partner of $k_a.$  The delta function in $x $ is a function of the
supertranslation invariant interval $x_a - x_a' - i \theta \sigma_a
\bar{\theta}' + i \theta' \sigma_a \bar{\theta}$ on  superspace\footnote{ Note
that $ \bar{{\cal D}}_{\dot{\alpha }} (x_a - x_a' - i \theta \sigma_a
\bar{\theta}' + i \theta' \sigma_a \bar{\theta}) = -i (\theta -
\theta')^{\alpha} (\sigma_a)_{\alpha \dot{\alpha}},$ so that although the
superspace invariant interval is not itself  left chiral, the fact that
$(\theta
- \theta')^3 = 0$ means that the full delta function on left chiral superspace
{\em is} annihilated by $ \bar{{\cal D}}_{\dot{\alpha}},$  as required.}.
Moving
the exponential to the left through the differential operators, the
coincidence
limit of the left chiral kernel has the expression
\begin{equation}
 K(t) =4 \int \frac{d^4k}{(2\pi)^4} \int d^2 \epsilon\,\, e^{t(X_aX_a +
W^{\alpha}X_{\alpha} + \frac12 ({\cal D}^{\alpha}W_{\alpha}))},
\label{chiralK}
\end{equation}
where
\begin{equation}
 X_a = {\cal D}_a + i k_a , \, \, \, \, \, X_{\alpha} = {\cal D}_{\alpha} + i
\epsilon_{\alpha}  .
\label{X}
\end{equation}
Note  that there is also a shift $ -  k_a (\sigma_a)_{\alpha \dot{\alpha}}
(\bar{\theta } - \bar{\theta}')^{\dot{\alpha}}$ in ${\cal D}_{\alpha}$;
however,
this vanishes in the coincidence limit as there are no $\bar{{\cal
D}}_{\dot{\alpha}}$ operators present to annihilate the $\bar{\theta} -
\bar{\theta}'.$
Also note that the integrand in (\ref{chiralK}) contains an explicit factor
$e^{-k^2}$ necessary for the convergence of the $k$ integral.
To compute the kernel, we will solve the differential equation
\begin{equation}
  \frac{dK(t)}{dt} = K_{aa}(t) + W^{\alpha} K_{\alpha}(t) + \frac12({\cal
D}^{\alpha}W_{\alpha}) K(t) ,
\label{dess}
\end{equation}
where
$$K_{A_1 A_2 \, \cdots A_n}(t) = 4 \int \frac{d^4k}{(2\pi)^4} \int d^2
\epsilon\,\,X_{A_1} X_{A_2} \, \cdots \, X_{A_n} \, e^{t\Delta},  $$
and $X_{A} $ can represent either a bosonic operator $X_a $ or a fermionic
operator $X_{\alpha}; $ the abbreviation  $\Delta = X_aX_a +
W^{\alpha}X_{\alpha} + \frac12 ({\cal D}^{\alpha}W_{\alpha}) $ has also been
introduced. The aim is to use identities similar to those used in \S2 to
express
$ K_{aa}(t) $ and $ W^{\alpha} K_{\alpha}(t) $ in terms of $K(t).$ In
superspace, there are two  kinds of identities involving the vanishing the
integral of a total derivative which can be employed, one involving
derivatives
with respect to the bosonic variables $k_a$ and the other involving
derivatives
with respect to the fermionic variables $\epsilon_{\alpha}$:
\begin{eqnarray}
 0 &=& \int \frac{d^4k}{(2\pi)^4} \int d^2
\epsilon\,\,\frac{\partial}{\partial
k^b} \left(X_{A_1} X_{A_2} \, \cdots \, X_{A_n} \, e^{t\Delta}
\right),\nonumber
\\
 0 &=& \int \frac{d^4k}{(2\pi)^4} \int d^2
\epsilon\,\,\frac{\partial}{\partial
\epsilon^{\beta}} \left(X_{A_1} X_{A_2} \, \cdots \, X_{A_n} \, e^{t\Delta}
\right).
\label{idents}
\end{eqnarray}
The expression for $K_{ab}(t)$ will arise from the use of  the first identity
with  $X_{A_1} X_{A_2} \, \cdots \, X_{A_n}$ replaced by $X_a,$ as to  the
action of the derivative on the exponential  pulls down an operator  which to
leading order is $2 i t X_a.$ Similarly, an expression for $W_{\beta}
K_{\alpha}(t) $ can be obtained from the second identity with  $X_{A_1}
X_{A_2}
\, \cdots \, X_{A_n}$ replaced by $X_{\alpha},$ as the action of the
derivative
on the exponential pulls down a factor which to leading order is $-i t
W_{\beta}.$

In both cases, the factors $2 i t X_a $ and  $-i t W_{\beta}$  will be
accompanied by additional terms containing commutators of $\Delta$ with these
factors. As was the case in \S2, these will form an infinite series which
cannot
be summed in general. It is necessary to truncate to a given order in
commutators and anticommutators of $X_{a}$ and $X_{\alpha},$ corresponding
to a
particular order in the (super)derivative expansion of the effective
action. To
lowest order, the effective action  contains  the superfields $W_{\alpha},$
but
no derivatives of them. Since $W_{\alpha}$ can be expressed   as a double
(anti)commutator of spinor derivatives ${\cal D}_{\alpha}$ and $\bar{{\cal
D}}_{\dot{\alpha}},$ this corresponds to truncation to at most  two
(anti)commutators of spinor derivatives  (with ${\cal D}_{a}$ counting as a
first order anticommutator via $ \{ {\cal D}_{\alpha}, \bar{{\cal
D}}_{\dot{\alpha}} \} =  -2 i (\sigma_a)_{\alpha \dot{\alpha} } \,{\cal
D}_a $).
The left chiral kernel is trivial to compute in this approximation, because
$[X_a,X_b],$ $[X_a, X_{\alpha}], $
 $[X_a, W_{\alpha}] $ and  $\{X_{\alpha}, W_{\beta}\} $ all involve at least
three (anti)commutators of  spinor derivatives and therefore must be set to
zero. The only potential anticommutator at this order is
$\{X_{\alpha},X_{\beta}\},$ but this vanishes due to the torsion and curvature
constraints imposed in supersymmetric Yang-Mills theory \cite{Wess}. The
kernel
thus reduces in this lowest order approximation to
\begin{equation}
K(t) =  4 \int \frac{d^4k}{(2\pi)^4} \int d^2 \epsilon\,\, e^{-tk^2}
e^{itW^{\alpha} \epsilon_{\alpha}} = \frac{1}{16 \pi^2} W^{\alpha}W_{\alpha}.
\label{K0}
\end{equation}
Note that this does not even reproduce the the nonsupersymmetric  low energy
effective actions considered in \S2; it contains only the leading term,
quadratic in the Yang-Mills field strength. To reproduce the supersymmetric
analogue of these results, it is necessary to go to the next order in the
superspace derivative expansion, namely to consider up to three
(anti)commutators of spinor derivatives. This is the truncation which will be
made here.

Using the curvature and torsion constraints for supersymmetric Yang-Mills
theory
\cite{Wess}, and letting $\bar{M}_{ab} = (\bar{{\cal D}} \bar{\sigma}_{ab}
\bar{W}), $ $M_{ab} = ({\cal D}\sigma_{ab}W), $ and $N_{\alpha \beta} = ({\cal
D}_{\alpha}W_{\beta}),$  the nonvanishing commutators in this truncation  are:
$$  [X_a,X_b] = -\frac12(\bar{M}_{ab} - M_{ab}),  \, \,  [X_a, X_{\alpha}] = i
(\sigma_a)_{\alpha \dot{\alpha}} \bar{W}^{\dot{\alpha}},\, \,  \{X_{\alpha},
W_{\beta}\} = N_{\alpha \beta}.$$ In particular, note that
$[X_a, W_{\alpha}]$ must be set to zero at  this order.

We first consider the calculation of $W^{\alpha}K_{\alpha}(t)$ in terms of
$K(t).$ The relevant identity is
\begin{eqnarray*}
 0 & =&  4 \int \frac{d^4k}{(2\pi)^4} \int d^2
\epsilon\,\,\frac{\partial}{\partial \epsilon_{\beta}} \left(X_{\alpha}  \,
e^{t\Delta} \right) \\
&=& i \delta_{\alpha}{}^{\beta} K(t) -  4 \int \frac{d^4k}{(2\pi)^4} \int d^2
\epsilon\,\, X_{\alpha}  \,\frac{\partial}{\partial \epsilon_{\beta}}
e^{t\Delta}.
\end{eqnarray*}
The derivative of the exponential is computed using
$$\frac{\partial}{\partial \epsilon_{\beta}} e^{t\Delta} =\left(  \int_0^1
ds \,
e^{st\Delta}\,  (-i t W^{\beta}) \, e^{-st \Delta}\right)  e^{t \Delta} = -i t
\sum_{n=0}^{\infty} \frac{t^n}{(n+1)!} ad^{(n)}(\Delta)(W^{\beta})
e^{t\Delta}.$$
The commutators  in the series are easily evaluated to the required order, and
we obtain
\begin{equation}
 \frac{\partial}{\partial \epsilon_{\beta}} e^{t\Delta} = -i t
\sum_{n=0}^{\infty} \frac{t^n}{(n+1)!} W^{\gamma} (N^n)_{\gamma }{}^{
\beta} = -
i  W^{\gamma}\left( \frac{e^{tN} - {\bf 1}}{N}\right)_{\gamma}{}^{\beta}.
\label{epsilonderiv}
\end{equation} Note that the power series expansion begins at order $N^0$ and
does not involve any negative powers of the matrix $N;$ thus the inverse
matrix
$ \left( \frac{N}{e^{tN} - 1} \right)_{\gamma}{}^{\beta} $ exists and does not
require the inversion of  $N.$ Substituting this result and being careful to
include the anticommutator which arises from moving $W^{\gamma}$ through
$X_{\alpha},$
we obtain
$$ 0 = \delta_{\alpha}{}^{\beta} K(t) + N_{\alpha}{}^{\gamma}\left(
\frac{e^{tN}
- {\bf 1}}{N}\right)_{\gamma}{}^{\beta} K(t) -  W^{\gamma}\left(
\frac{e^{tN} -
{\bf 1}}{N}\right)_{\gamma}{}^{\beta} K_{\alpha}(t). $$
Multiplying by $ \left( \frac{N}{e^{tN} - {\bf 1}} \right)_{\beta}{}^{\rho} $
and contracting indices appropriately yields a result of the desired form:
\begin{equation}
W^{\alpha}K_{\alpha}(t) = tr  \left( \frac{N}{e^{tN} - \bf{1}} \right)\,
K(t) +
tr(N) \, K(t).
\label{WK}
\end{equation}
Note that the trace here is over the spinor indices, $tr(N) =
N_{\alpha}{}^{\alpha};$ there is  {\em not} a trace over gauge indices, as the
expression is still a matrix with respect to gauge indices.

The first identity in (\ref{idents}) is used to compute $K_{aa}(t)$ in the
same
manner as in the bosonic  case in \S2.
To the order in which we are working, one finds
$$ \frac{\partial}{\partial k^b}  e^{t \Delta}  =
2 i t B_{bc}(t) X_c + 2 i t A_b(t),$$
where
\begin{eqnarray*}
 B_{bc}(t) & = &  \left( \frac{e^{-t(M-\bar{M})} - {\bf 1}}{-t(M-\bar{M})}
\right)_{bc}\\
A_b(t) & = & \frac{i}{t} \biggl[\frac{(e^{-t(M-\bar{M})} - {\bf 1})}{(M-
\bar{M}) (\bar{M} - \frac12 tr(N) {\bf 1}) }\\
&  - &  \frac{(e^{-t(M-\frac12 tr(N) {\bf 1})}-{\bf 1})}{ (M -  \frac12 tr(N){
\bf 1}) (\bar{M} - \frac12 tr(N) {\bf 1})} \biggr]_{bc}\, (W^{\alpha}\sigma_{c
\alpha \dot{\alpha}} \bar{W}^{\dot{\alpha}}).
\end{eqnarray*}
Thus  the first  identity in (\ref{idents}) with $X_{A_1} X_{A_2} \, \cdots \,
X_{A_n}$ replaced by $X_{a}$  yields
$$ 0 = i \delta_{ab} K(t)+ 2 i t B_{bc}(t) K_{ac}(t) + 2 i t A_b(t) K_{a}(t).$$
The matrix $B_{bc}(t)$ has a power series expansion in positive powers of $M$
and $\bar{M}$ which begins at order ${\bf1},$ and so it is invertible, yielding
$$ K_{ab}(t) = -\frac{1}{2t} (B^{-1})_{ba}(t) K(t) -  (B^{-1})_{bc}(t) A_c(t)
K_a(t).$$
The right hand side involves $K_{a}(t); $ this is evaluated by  using the
first
identity in identity (\ref{idents}) with $ X_{A_1} X_{A_2} \, \cdots \,
X_{A_n}
$ replaced by $1.$ The result is
\begin{equation}
 K_a(t) = - (B^{-1})_{ab}(t) A_b(t) K(t).
\label{Ka}
\end{equation}

At this point an important simplification can be achieved by noting that the
kernel in the order that we currently working must reduce to the ``zero'th
order'' result $K(t) = \frac{1}{16\pi^2}\,  W^2$ as $M,\bar{M}$ and $ N
\rightarrow 0$ i.e. when the higher order commutators we have allowed at this
order vanish. So the kernel must be of the form
$K(t) = F[M,\bar{M},N] \, \frac{1}{16\pi^2} W^2$ with $ F[M,\bar{M},N]
\rightarrow 1$ as  $M,\bar{M}, N \rightarrow 0.$ On the other hand, to the
order
that we are working, $\{W_{\alpha}, W_{\beta}\}$ can be taken to be zero,
since
it involves a five commutators of spinor derivatives\footnote{In a nonabelian
gauge theory, it is not true in general that  $\{W_{\alpha}, W_{\beta}\}= 0;$
rather,  $\{W_{\alpha}, W_{\beta}\} = W_{\alpha}^a W_{\beta}^b f_{ab}{}^c
T_c,$
where $T_a$ are generators of the gauge group satisfying $[T_a,T_b] =
f_{ab}{}^cT_c.$  Also note that using the general result
$W_{\alpha} W_{\beta} = \frac12 \epsilon_{\alpha \beta} W^2 + \frac12 \{
W_{\alpha}, W_{\beta} \},$
it follows that to the order we are working, $W_{\alpha} W_{\beta} = \frac12
\epsilon_{\alpha \beta} W^2.$ }. As a result, expressions involving more than
two $W's$ vanish at the order we are working because one index must be
repeated
and $(W_{\alpha})^2 = 0.$ Therefore, using (\ref{Ka}),  $K_a(t)$ vanishes
because $A_b(t)$ involves one factor of $W$ and $K(t)$ involves two. Thus the
expression for $K_{ab}(t)$ becomes simply:
\begin{equation}
 K_{ab}(t) = -\frac{1}{2t} (B^{-1})_{ba}(t) K(t).  \label{Kabss}
\end{equation}

Substituting (\ref{WK}) and (\ref{Kabss})  into (\ref{dess}), the differential
equation for the  left chiral kernel is
$$ \frac{dK(t)}{dt} =\frac12  tr\left( \frac{(M-
\bar{M})}{e^{-t(M-\bar{M})}-{\bf 1}} \right)\, K(t) + tr\left(
\frac{N}{e^{tN}-{\bf 1}}\right) \, K(t) +  \frac12 tr(N) \, K(t).$$
Noting the similarity of the first and second terms on the right-hand side of
the equation with the differential equation in \S2, the solution is easily
seen
to be
$$K(t) = c_1 \,e^{\frac{t}{2} tr(N)} \, \det \left(\frac{{\bf 1} - e^{-t
N}}{A_1} \right) \, \det \left(\frac{{\bf 1} - e^{t(M-
\bar{M})}}{A_2}\right)^{-\frac12},$$
where $c_1, A_1$ and $A_2$ are constants (independent of $t$). The latter are
determined by the requirement that in the limit $M, \bar{M}, N \rightarrow 0,$
the kernel must be of the form (\ref{K0}), from which it follows that
\begin{equation} K(t) = \frac{W^2}{16 \pi^2} \,\, e^{\frac{t}{2} tr(N)} \,
\det
\left(\frac{{\bf 1} - e^{-t N}}{N} \right) \, \det \left(\frac{{\bf 1} -
e^{t(M-
\bar{M})}}{(M- \bar{M}) }\right)^{-\frac12}.
\label{kernelresult}
\end{equation}
The power series expansion of the chiral kernel begins at order $t^0,$ and
so is
of the form $\sum_{n=0}^{\infty} V_n t^n.$ This yields the
 vertices in the effective action for the Yang Mills superfield background via
(\ref{seriesexp}). As mentioned in the introduction, this case is
not considered in the recent computations of effective actions for
supersymmetric Yang-Mills theories by graphical techniques
\cite{dewit,grisaru}, or in earlier results of
Buchbinder et al \cite{buchbinder1,buchbinder2} using functional
methods. Pickering and West \cite{pickering} have computed the piece
of the effective action corresponding to the lowest order
approximation (\ref{K0}) to the heat kernel, but have not computed the
corrections involving $ \cal{D}_{\alpha}W_{\beta}.$

We can perform two checks on the result (\ref{kernelresult}). The first is
to go back to the
expression (\ref{chiralK}) and perform the $\epsilon$ integral explicitly.
This
is done  using  $ 4\int d^2 \epsilon = \frac{\partial}{\partial
\epsilon_{\alpha}} \frac{\partial}{\partial \epsilon^{\alpha}}|_{\epsilon =
0}$
and (\ref{epsilonderiv}). One finds
\begin{eqnarray*}
K(t) &=&  \frac12 \, W^2\,  tr \left[ \left( \frac{e^{tN}-{\bf 1}}{N}
\right) \,
\left( \frac{e^{-t(N - tr(N) {\bf 1})}-{\bf 1}}{N- tr(N) {\bf 1}}\right)
\right]
\\
& & \int \frac{d^4k}{(2\pi)^4} \, \exp \left\{ t(X_a X_a + W^{\alpha}{\cal
D}_{\alpha}  + \frac12 (D^{\alpha}W_{\alpha}))\right\}.
\end{eqnarray*}
Note that since $\epsilon$ has been set to zero, the integral involves the
operator $W^{\alpha}{\cal D}_{\alpha} $ rather than  $W^{\alpha} X_{\alpha}.$
The remaining  integral can be evaluated by the trick of differentiating with
respect to $t$ and using total $k$ derivative identities. The result is
\begin{eqnarray*}
K(t) & =  &  -\frac{1}{32\pi^2} \,  W^2 \, tr \left[ \left( \frac{e^{tN}-{\bf
1}}{N} \right) \, \left( \frac{e^{-t(N - tr(N) {\bf 1})}-{\bf 1}}{N- tr(N)
{\bf
1}}\right) \right]\, e^{-\frac{t}{2} tr N}\\
& &  \det \left(\frac{{\bf 1}-e^{t(M-\bar{M})}}{(M-\bar{M})}
\right)^{-\frac12}.
\end{eqnarray*}
Although this seems to differ from (\ref{kernelresult}), in fact the two
expressions can be shown to be equivalent. This relies on  the fact that  $N$
 is a $2\times 2$ matrix, so that $\det N = \frac12 (trN)^2 - \frac12
tr(N^2).
$ Also, $tr \left(e^{-tN}\right) $ $= e^{-\frac{t}{2} tr(N)} \, tr \,
\left[e^{-t(N-\frac12 tr(N) {\bf 1})}\right],$ and since $N-\frac12 tr(N) {\bf
1}$ is a traceless $2\times 2$ matrix, only traces of even powers are
nonvanishing, with $tr[(N-\frac12 tr(N) {\bf 1})^{2n}] = 2 [\frac12
tr(N-\frac12
tr(N) {\bf 1})^2]^n.$

The second check is to compare the result (\ref{kernelresult}) with component
results. In the case where the fermionic components of the supersymmetric
Yang-Mills background are set to zero, we expect that the supersymmetric
kernel
should reduce to the difference of the bosonic kernels in \S2 for spin zero
and
spin half quantum fields in a  Yang-Mills background. To see that this is so,
consider a supersymmetric theory  in the presence of a supersymmetric
background. The one loop effective action for the background fields is
minus the
 logarithm of the  superdeterminant of the operator appearing in the part
of the
action quadratic in the quantum fields.  If the fermionic components of the
background fields are set to zero, then this superdeterminant factorises
into a
ratio of ordinary determinants, so
$$ \Gamma_{eff} = \ln \det \Delta_B -  \ln \det \Delta_{F}, $$
where $ \Delta_B $ and  $  \Delta_{F}$ denote the operators in the pieces
of the
action quadratic in the bosonic and fermionic quantum fields respectively.
This
is equivalent to $-\zeta^{\prime}_{B}(0) + \zeta^{\prime}_{F}(0),$ where
$\zeta_{B}(s) $ and $ \zeta_{F}(s) $are the usual zeta functions associated
with
the operators  $ \Delta_B $ and  $  \Delta_{F}$ respectively.  The
difference $
\zeta_{B}(s) -  \zeta_{F}(s)$ of the zeta functions is of the form
$$\frac{1}{\Gamma(s)} \int_0^{\infty} ds \, t^{s-1} \int d^4x \, K(t/\mu^2), $$
where
$$ K(t) = \lim_{x \rightarrow x'} \, ( e^{t\Delta_B} - e^{t \Delta_F}) \,
\delta^{(4)}(x,x') \equiv K_B(t) - K_F(t). $$
As the latter is  difference of functional traces, it is  a functional {\em
supertrace}.

In the case at hand, if the fermionic components of the background Yang-Mills
superfields are set to zero, then, with the convention $[D_a,D_b] = F_{ab},$
$ N_{\alpha}{}^{\beta}   \rightarrow   F_{ab}
(\sigma_{ab})_{\alpha}{}^{\beta},
$ $ Tr(N) \rightarrow 0, $ $  (M - \bar{M})_{ab} \rightarrow  -2 F_{ab}, $
and $  W_{\alpha} \rightarrow - F_{ab} (\sigma_{ab})_{\alpha}{}^{\beta}
\theta_{\beta}. $
Using the antisymmetry of $F_{ab},$ the supersymmetric kernel
(\ref{kernelresult}) becomes
$$K(t) = - \theta^2   \det \left({\bf 1}-e^{-tF}\right)  \frac{1}{4\pi^2} \det
\left(\frac{{\bf 1}- e^{2tF}}{F}\right)^{-\frac12}$$
where the first determinant is over $2\times   2$ undotted spinor indices with
$F \equiv F_{ab} (\sigma_{ab})_{\alpha}{}^{\beta}$, and the second determinant
is over vector indices with  $F \equiv F_{ab}$. On the other hand, using
(\ref{le}) and (\ref{spin12}), the difference of the kernel for a scalar and
for a left-handed spinor\footnote{The normalization of the left handed spinor
kernel is  changed by a factor of $\frac12$ relative to that of the scalar
kernel because the kernel (\ref{spin12}) is computed for the {\em square}
of the
Dirac operator, whereas the effective action contains  the determinant of the
Dirac operator and not its square.}
$$  \frac12  tr \left( {\bf 1}- e^{-tF}\right)  \frac{1}{4\pi^2} \det
\left(\frac{{\bf 1}- e^{2tF}}{F}\right)^{-\frac12},$$
where again the trace is over $2\times 2$ spinor indices and the
determinant is
over vector indices. Using the fact that $ F_{ab}
(\sigma_{ab})_{\alpha}{}^{\beta}$ is a traceless $2\times 2$ matrix, it is
relatively easy to show the equivalence of  $- \det \left({\bf
1}-e^{-tF}\right)$ and  $tr\left( {\bf 1}- e^{-tF}\right),$ thus ensuring
that
the supersymmetric kernel (\ref{kernelresult}) for a purely  bosonic
background
does indeed reduce to a difference of kernels  for bosonic and fermionic
quantum
fields.

\section{ Discussion}
In this paper, we have illustrated a new approach to computing effective
actions
in supersymmetric theories by applying  it to the determination of
 the low energy effective action obtained by
integrating out massive scalar supermultiplets in the presence of a
supersymmetric Yang-Mills background. The approach relies on computing the
functional trace of an appropriate heat kernel; however, it is not
necessary to
compute Green's functions for the quantum fields in the presence of the
background, as  advocated  by Buchbinder et al \cite{buchbinder1,buchbinder2}.
Instead, we make use of the similarity of the functional trace of the heat
kernel expressed in momentum space  to a Gaussian integral. It is expected
that
the method will also be applicable in the case of scalar superfield
backgrounds.

The mass dependence of the effective action involved the Mellin transform of a
generalized hypergeometric function. Although we were unable to evaluate  the
Mellin transform explicitly, the fact that the hypergeometric function reduced
to an exponential in the limit $s \rightarrow 0$ was enough to determine the
form of the mass dependence of the effective action. We note in passing
that in
the calculation of Buchbinder et al \cite{buchbinder1} of the low energy
effective K\"ahler potential for the N=1 Wess-Zumino model, the kernel also
involves a generalized hypergeometric function.
 The functional supertrace of the appropriate heat kernel is \cite{buchbinder1}
\begin{equation}
K(t) = \frac{\eta \bar{\eta}}{16 \pi^2} \sum_{n=1}^{\infty}
\frac{n!}{(2n)!} (-t
\eta \bar{\eta})^{n-1}.
\label{hyp}
\end{equation}
Here, $\eta$ is the left chiral superfield $ \frac{\partial^2 W}{\partial \Phi
\partial \Phi},$ where $W(\Phi)$ is the superpotential for the chiral scalar
superfields $\Phi$ in the Wess-Zumino model.
The sum on the right hand side is actually  a generalized hypergeometric
function,
$$K(t) = \frac{\eta \bar{\eta}}{32\pi^2}\,  {}_1F_1[1,\frac32 ; -\frac{t \eta
\bar{\eta}}{4}].$$
This simplifies considerably the computation of the zeta function
$$ \zeta(s) =   \frac{1}{\Gamma(s)} \int_0^{\infty} dt \, t^{s-1}\,  \int d^8Z
\,\,  K(t/\mu^2).$$
Using the integral representation
$$ {}_1F_1[\alpha, \gamma; z] = \frac{\Gamma(\gamma)}{\Gamma(\alpha)
\Gamma(\gamma -\alpha)} \int_0^1 dx \, e^{zx} \, x^{\alpha -1} (1-x)^{\gamma -
\alpha -1} $$
for the confluent hypergeometric functions,
\begin{eqnarray*}
 \zeta(s) &=& \int d^8Z \,  \frac{\eta \bar{\eta}}{64\pi^2 \Gamma(s)} \,
\, \,
\int_0^1 dx \, (1-x)^{-\frac12} \, \int_0^{\infty} dt  \, t^{s-1} \,
e^{-\frac{t
\eta \bar{\eta x }}{4 \mu^2}} \\
&=&\int d^8Z \,  \frac{\eta \bar{\eta}}{64\pi^2 } \, \left (\frac{4
\mu^2}{\eta
\bar{\eta}} \right)^s \int_0^1 dx \, \frac{(1-x)^{-\frac12}}{x^s}.
\end{eqnarray*}
The integral over $x $ is just the beta function $B(\frac12, 1-s) =
\frac{\Gamma(\frac12)\Gamma(1-s)}{\Gamma(\frac32 -s)}.$
>From this it is easy to show
$$\zeta^{\prime}(0) =\int d^8Z \, \frac{\eta \bar{\eta}}{32\pi^2 }\,  (2 -
\ln
\frac{\eta \bar{\eta}}{\mu^2}),$$
which is the result obtained by Pickering and West  in \cite{pickering} using
supergraph techniques. The Schwinger proper time regularization employed by
Buchbinder et al \cite{buchbinder1} yields an expression for the effective
K\"ahler potential which  contains a finite constant  in the form of  an
infinite series,  which is not evaluated explicitly. Thus recognizing that the
kernel is essentially a hypergeometric function in this case yields
considerable
simplification in the computation of the effective action if zeta function
regularization is employed.

Finally, we make a comment of the form of the effective action
(\ref{kernelresult}). The effective action computed using the lowest
order approximation (\ref{K0}) to the heat kernel is holomorphic in
the sense of Seiberg \cite{Seiberg}, in that it contains only terms
with chiral or antichiral gauge superfields. However, when
corrections incorporating $\cal{D}_{\alpha}W_{\beta}$ are included as
in (\ref{kernelresult}), this is no longer true, and the result
contains both $\bar{M}_{ab} = (\bar{{\cal D}} \bar{\sigma}_{ab}
\bar{W}) $  and $M_{ab} = ({\cal D}\sigma_{ab}W)$ in the one term.
 In calculating the
effective action for N=2 supersymmetric Yang-Mills theory in N=1
superfield formulation, the result (\ref{kernelresult}) will thus represents
a nonholomorphic correction to the effective action which should be
included with the nonholomorphic corrections involving scalar
superfields computed in \cite{dewit,pickering,grisaru}.

\ack{ IMcA is grateful to the Alexander von Humboldt-Stiftung for support.
Both
authors wish to thank Professor J. Wess for his generous hospitality at the
University of M\"unchen during the period in which this  this work was done.}

\end{document}